\documentclass[conference, 10pt]{IEEEtran}
\usepackage{bbm}
\usepackage{bbold}
\usepackage{amsmath}
\usepackage{amssymb}
\usepackage{breqn}
\usepackage[dvips]{graphicx}
\usepackage{graphicx}
\usepackage{subfigure}
\usepackage{epsfig} 
\usepackage{bigints}
\usepackage{stfloats}
\usepackage{amsthm}
\usepackage{eso-pic}

\newtheorem{prop}{Proposition}
\newtheorem{lemma}{Lemma}
\newtheorem{corollary}{Corollary}

\ifCLASSINFOpdf
\else
\fi

\IEEEoverridecommandlockouts 
\IEEEpubid{\makebox[\columnwidth]{978-1-4673-6540-6/15/\$31.00~\copyright~2015 IEEE \hfill} \hspace{\columnsep}\makebox[\columnwidth]{ }}

\begin{document}
%
% paper title
% can use linebreaks \\ within to get better formatting as desired
\title{Modeling and Analysis of Content Caching in Wireless Small Cell Networks
}
% author names and affiliations
% use a multiple column layout for up to three different
% affiliations
\author{\IEEEauthorblockN{Syed Tamoor-ul-Hassan\IEEEauthorrefmark{1}, Mehdi Bennis\IEEEauthorrefmark{1}, Pedro H. J. Nardelli\IEEEauthorrefmark{1}, Matti Latva-Aho\IEEEauthorrefmark{1}}
\thanks{This research has been supported by Finnish Funding Agency for Technology and Innovation (TEKES) under the project Multi-Operator Spectrum Sharing for Future 5G Networks (MOSSAF) (Award 2349/31/2013), and Academy of Finland.
}
\IEEEauthorblockA{\IEEEauthorrefmark{1}Center for Wireless Communications, University of Oulu, Finland,\\
Email: \{tsyed, bennis, nardelli, matti.latva-aho\}@ee.oulu.fi}
}
 
\providecommand{\keywords}[1]{\textbf{\textit{Index terms---}} #1}

%%\and
%\IEEEauthorblockN{Mehdi Bennis}
%\IEEEauthorblockA{Center for Wireless Communications\\
%University of Oulu\\
%Finland \\
%Email: bennis@ee.oulu.fi}
%\and
%\IEEEauthorblockN{Matti Latva-Aho}
%\IEEEauthorblockA{Center for Wireless Communications\\
%University of Oulu\\
%Finland \\
%Email: matti.latva-aho@ee.oulu.fi}

% conference papers do not typically use \thanks and this command
% is locked out in conference mode. If really needed, such as for
% the acknowledgment of grants, issue a \IEEEoverridecommandlockouts
% after \documentclass

% for over three affiliations, or if they all won't fit within the width
% of the page, use this alternative format:
% 
%\author{\IEEEauthorblockN{Michael Shell\IEEEauthorrefmark{1},
%Homer Simpson\IEEEauthorrefmark{2},
%James Kirk\IEEEauthorrefmark{3}, 
%Montgomery Scott\IEEEauthorrefmark{3} and
%Eldon Tyrell\IEEEauthorrefmark{4}}
%\IEEEauthorblockA{\IEEEauthorrefmark{1}School of Electrical and Computer Engineering\\
%Georgia Institute of Technology,
%Atlanta, Georgia 30332--0250\\ Email: see http://www.michaelshell.org/contact.html}
%\IEEEauthorblockA{\IEEEauthorrefmark{2}Twentieth Century Fox, Springfield, USA\\
%Email: homer@thesimpsons.com}
%\IEEEauthorblockA{\IEEEauthorrefmark{3}Starfleet Academy, San Francisco, California 96678-2391\\
%Telephone: (800) 555--1212, Fax: (888) 555--1212}
%\IEEEauthorblockA{\IEEEauthorrefmark{4}Tyrell Inc., 123 Replicant Street, Los Angeles, California 90210--4321}}

% use for special paper notices
%\IEEEspecialpapernotice{(Invited Paper)}

% make the title area
\maketitle

\begin{abstract}
%\boldmath
Network densification with small cell base stations is a promising solution to satisfy future data traffic demands. However, increasing small cell base station density alone does not ensure better users' quality-of-experience and incurs high operational expenditures. Therefore, content caching on different network elements has been proposed as a mean of offloading the backhaul  by caching strategic contents at the network edge, thereby reducing latency. In this paper, we investigate cache-enabled small cells in which we model and characterize the outage probability, defined as the probability of not satisfying users' requests over a given coverage area. We analytically derive a closed form expression of the outage probability as a function of signal-to-interference ratio, cache size, small cell base station density and threshold distance. By assuming the distribution of base stations as a Poisson point process, we derive the probability of finding a specific content within a threshold distance and the optimal small cell base station density that achieves a given target cache hit probability. Furthermore, simulation results are performed to validate the analytical model.     
% The predicted data traffic target cannot be improved despite the network densification by Small Cell Base Stations (SCBA). Caching popular contents on different network elements significantly improves the network performance in terms of backhaul offloading, network cogestion and users QoE. In this paper, we analytically model   
\end{abstract}
%% IEEEtran.cls defaults to using nonbold math in the Abstract.
%% This preserves the distinction between vectors and scalars. However,
%% if the conference you are submitting to favors bold math in the abstract,
%% then you can use LaTeX's standard command \boldmath at the very start
%% of the abstract to achieve this. Many IEEE journals/conferences frown on
%% math in the abstract anyway.
%
%% no keywords
%
%
%
%
%% For peer review papers, you can put extra information on the cover
%% page as needed:
%% \ifCLASSOPTIONpeerreview
%% \begin{center} \bfseries EDICS Category: 3-BBND \end{center}
%% \fi
%%
%% For peerreview papers, this IEEEtran command inserts a page break and
%% creates the second title. It will be ignored for other modes.
%\IEEEpeerreviewmaketitle
%
%
%
%\section{Introduction}
%% no \IEEEPARstart
%This demo file is intended to serve as a ``starter file''
%for IEEE conference papers produced under \LaTeX\ using
%IEEEtran.cls version 1.7 and later.
%% You must have at least 2 lines in the paragraph with the drop letter
%% (should never be an issue)
%I wish you the best of success.
%
%\hfill mds
 %
%\hfill January 11, 2007
%\AddToShipoutPicture*{\small \raisebox{1.5cm}{\hspace{1.5cm} \ 978-1-4673-6540-6/15/\$31.00 \copyright2015 IEEE}}
\section{Introduction}
\label{sec:Intro}
The demand for mobile broadband is increasing rapidly with emerging applications and services \cite{Ref1}. These services require an optimized quality-of-service (QoS) for efficient network operation. Current state-of-the-art research focuses on traditional techniques such as interference management, cooperative communication and cognitive radio to improve spectrum utilization \cite{Ref2:TrdRef1}. However, these techniques cannot meet the predicted data traffic growth alone without additional spectrum and more pronounced network densification \cite{Ref4:Energy2}. Therefore, heterogeneous networks, comprising of small cell base stations (SBSs) underlying the macrocellular network, constitute a promising solution to improve coverage and boost capacity \cite{Ref3:SCN_2} \cite{Ref3:SCN_3}. However, simply adding small cells incurs high capital and operational expenditures, which limits their deployments. To remedy to this, content caching at the network edge has recently been identified as one of the five most disruptive paradigms in 5G networks \cite{Ref:Boc_5G}. Dynamic caching can significantly offload different parts of the network including the radio access network, core network, and backhaul, by smartly prefetching and storing contents closer to the end-users. As a result, network congestion is eased, backhaul is offloaded and users' quality-of-experience (QoE) is maximized \cite{Ref6:Intro1}. \\ 
\indent A significant work has been done on content caching in wireless networks. %A significant work has been done on performance evaluation of caching \cite{Ref7:RelWrk1} \cite{Ref7:RelWrk2} and its applications for various wired networks i.e., Content Distribution Networks (CDNs) \cite{Ref7:RelWrk3} \cite{Ref7:RelWrk4} and wireless networks \cite{Ref7:RelWrk5}-\cite{Ref7:RelWrk12}.
In \cite{Ref7:RelWrk5}, the author presented and optimized opportunistic cooperative MIMO (CoMP) scheme for wireless video streaming by caching part of video files at the relays. 
%and optimize the scheme by creating CoMP opportunities followed by dynamic power control. 
%Proactive resource allocation is suggested in \cite{Ref7:RelWrk6}, where user behavior is predicted to balance wireless traffic, leading to improved resource utilization for target outage probability. 
In \cite{Ref7:RelWrk7}, data is cached on different wireless caches with limited storage capabilities and the performance of uncoded and coded data transmission is evaluated by minimizing the distance to retrieve the complete data file. Collaborative caching is presented in \cite{Ref7:RelWrk10} where the social welfare is maximized using Vickrey-Clarke-Groves (VCG) auctions. 
%The work also investigates the impact of collaborative caching on the quality of video streaming. 
Femtocaching was proposed in %\cite{Ref7:RelWrk9} 
\cite{Ref7:RelWrk8} where caching helpers store popular contents optimally, thus improving frequency reuse. 
%With the proposed distributed network, the number of satisfied users are increased by 600-700\% in \cite{Ref7:RelWrk8}. 
\cite{Ref6:Intro1} examined proactive caching by prefetching contents on different network elements, thus
%. The proactive caching paradigm provides 
providing substantial gains in terms of backhaul savings and satisfied users. 
%The setup in \cite{Ref6:Intro2} leverages storage capabilities to offload backhaul link, thus, increasing the satisfaction level of UEs. 
%Similarly, in \cite{Ref6:Intro3}, the author shows the decrease in delay by content localization. The author in \cite{Ref7:RelWrk8} shows the benefit of content localization in terms of frequency reuse and guaranteed QoS.  
\cite{Ref7:RelWrk12} evaluates the performance of caching on helper stations/devices and optimize video quality by devising optimal storage schemes. \cite{Ref8:Caching_Ref1} estimates the loss rate in content delivery networks and proposes several replication methods. 
%and dynamic streaming techniques. 
%The network throughput of the proposed methods are compared with the traditional schemes.
Most of the works above consider fixed location of caches. To the best of our knowledge, \cite{Ref8:Caching_Ref2} was the first work  to study stochastic geometric approaches to caching. However, this work didn't address the optimization of SBS density as a function of cache size and distance. \cite{Ref7:RelWrk7} also considers content caching from a stochastic geometry perspective. However, instead of link reliability, it considers the distance to retrieve all parts of the requested file, without taking interference into account. \\ 
%Therefore, unlink \cite{Ref8:Caching_Ref1} and \cite{Ref7:RelWrk7}, 
%With the exception of \cite{Ref8:Caching_Ref2}, as per our knowledge, none of the above work models the cache miss probability in terms of SINR, SBS density and cache size. \\
\indent Unlike \cite{Ref8:Caching_Ref2} and \cite{Ref7:RelWrk7}, this paper investigates the problem of content caching in dense Small Cell Networks (SCNs), from a stochastic geometry perspective. SBSs are randomly deployed in the network according to a Poisson point process (PPP), and store a limited number of contents, from a library. By assuming a single user located at the origin of the network, 
%(i.e., \textit{typical user})
we derive a closed form expression of the outage probability, as a function of the SBS density, storage size and threshold distance. The outage probability is defined as the probability of accessing a specific content, cached in a given coverage area, subject to a given signal-to-interference (SIR)-dependent coverage probability. Moreover, for a given target cache hit probability, we derive a closed form expression of the outage probability assuming the closest SBS is within the area defined by threshold distance. By assuming a given threshold distance and cache size, we characterize the optimum number of SBSs required to achieve a target outage probability, and derive a number of insights. \\     
\indent The remainder of this paper is organized as follows: Section \ref{sec:Sys_Mod} describes the system model where the network and channel models are presented. In Section \ref{sec:Out_Pro_Form}, the content outage probability  is formalized and the optimum SBS density is characterized, for a fixed replication ratio and threshold distance to achieve a target cache hit probability. Finally, numerical results are presented in Section \ref{sec:Num_Res_Dis} along with simulation results.
\begin{figure}[t]%
\centering%
\includegraphics[scale = 0.55]{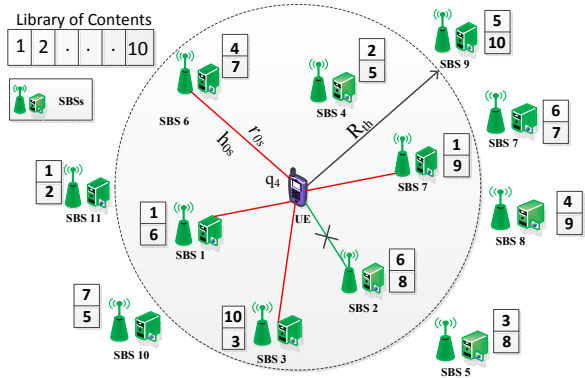}
\caption{Illustration of the network deployment under consideration. UE's content request $q_4$ is served by SBS $6$ within threshold distance $R_{\mathrm{th}}$.}
\label{fig:Net_Diagram1}
\end{figure}%
\vspace{-0.3cm} 
\section{System Model}
\label{sec:Sys_Mod}
%\subsection{Notations}
%\label{ssec:notation}
%The double stoke letters represent the probability of an event i.e., $\mathbbm{E}$. With the exception of $\mathbbm{\epsilon}$ which represents target cache hit probability, the probability and expectation of an event is represented by $\mathbbm{P}$ and $\mathbbm{E}$ respectively.  
\subsection{Network Model}
\label{ssec:Net_Mod}
\vspace{-0.1cm}
\indent Consider the downlink transmission of a wireless small cell network comprising of SBSs in a two dimensional Euclidean plane $\mathbb{R}^2$, as shown in Fig. \ref{fig:Net_Diagram1}. The distribution of SBSs is modeled as a homogeneous PPP, $\Phi_{\mathrm{s}}$
%$\Phi_s = \{X_s\}$, where $X_s$ represents the spatial location of $s$-th SBS
with intensity $\lambda_{\mathrm{s}}$ where each SBS is equipped with a cache of size $d$, to cache contents from a given library $\mathcal{C} = \{1,2, ..., |\mathcal{C}|\}$  with size $|\mathcal{C}|$. 
%Let $\Omega_{s}$ represents the contents cached at the $s$-th SBS. 
All contents are assumed to be of the same size. We assume the content popularity to follow a uniform distribution where each content is equally popular. As a result, each SBS randomly selects and caches contents from the library. \vspace{-0.2cm}
\subsection{User Association and Channel Modeling}
\vspace{-0.1cm}
\indent Without loss of generality, we consider a single user equipment (UE) located at the origin of the network, termed as a \textit{reference user} (also known as \textit{tagged user}). Such reference user has a threshold distance $R_{\mathrm{th}}$ that specifies the maximum distance over which its content request can be served. The reference user is associated to the nearest SBS with the cached content within $R_{\mathrm{th}}$. If no SBS caches the requested content, an outage event occurs. To simplify the mathematical analysis, we further consider that the original process $\Phi_{\mathrm{s}}$ contributes only to the interference field since the reference link is not part of the original PPPs. For this reason, the reference link emulates the user association process by assuming that the distance between the reference SBS and the reference user follows the distribution of the closest SBS having the cached content, conditioned on its existence. \\
\indent The standard power loss propagation channel model is used with path loss exponent $\alpha > 2$. In order to account for the random channel effects such as fading and shadowing, we consider a Rayleigh fading channel in which the channel gain from SBS $s$ to a reference user is $h_{0s}$, which is an exponential random variable with mean 1. We consider a constant transmit power per SBS. Moreover, we assume an interference-limited regime and neglect the effect of Additive White Gaussian Noise (AWGN). The SIR from SBS $s$ to the reference user is given as:
%\begin{equation*}
%\text{SINR}_{b}(r) = \frac {h_{b}r_{b}^{-\alpha}} {\sum_{\forall b' \in \mathcal{B}/b} h_{b'}r_{b'}^{-\alpha} + N_0}
%\end{equation*}
\begin{equation}
%\text{SIR}(r_b) = \frac {h_{s}r_{s}^{-\alpha}} {\sum_{\forall s' \in \Phi_{\mathbb{s}}} h_{s'}r_{s'}^{-\alpha} }, 
\text{SIR} = \frac {h_{0s}r_{0s}^{-\alpha}} {\sum_{\forall s' \in \Phi_{\mathrm{s}}} h_{0s'}r_{0s'}^{-\alpha} }, 
\end{equation}
where $r_{0s}$ is the random distance between the reference UE and SBS $s$ and $\alpha$ is the path loss exponent. \\
\indent The performance of the system depends on several factors such as $\lambda_{\mathrm{s}}$, $R_{\mathrm{th}}$ and $d$. A high SBS density increases the cache hit probability at the cost of an increased interference 
%which decreases the downlink transmission rate
. Meanwhile, increasing the threshold distance increases the coverage area of a given user, leading to a high cache hit. However, the SIR decreases with the increased distance (assuming free space path loss). Additionally, a large cache size increases the cache hit without affecting the interference for the same number of SBSs. The goal of the work is to maximize the cache hit probability while achieving a pre-defined SIR threshold based on the system parameters such as $\lambda_{\mathrm{s}}$, $R_{\mathrm{th}}$ and $d$. In the following section, the analytical results that capture such tradeoffs are presented.
\vspace{-0.2cm} 
\section{Outage Probability}
\label{sec:Out_Pro_Form}
In this section, after defining the coverage probability and cache hit probability, we optimize the SBS density to achieve a target cache hit probability based on the threshold distance and replication ratio. In addition, we derive the outage probability of getting a specific content. 
%\subsection{Definitions}
\vspace{-0.2cm}
\subsection{Replication Ratio}
The replication ratio, denoted by $P_c$, is the fraction of the library contents stored by the SBS. In this paper, we assume a uniform content replication so that the replication ratio is the same for every content $c$ and it is given by:
\vspace{-0.1cm}
\begin{equation}
P_c = \frac {d} {|\mathcal{C}|}.
\vspace{-0.2cm}
\end{equation}
Note that, as far as the uniform replication is assumed, the ratio $P_c$ is equal to the probability of having a given content $c$ in the SBS cache, with size $d$.
%The replication ratio, denoted by $P_c$, is the fraction of the library contents stored by the SBS. For a uniform content replication, the replication ratio is the same as the probability of the existence of the content and is given by:
%If the library size is $|\mathcal{C}|$ and the SBSs have a cache size $d$, then the replication ratio, assuming a uniform content replication, is given by:
\vspace{-0.2cm}
\subsection{Outage Probability}
The outage probability in an interference-limited system is defined as the probability that a reference UE does not achieve an SIR threshold ($\gamma$). Mathematically, the outage probability, with respect to an SBS $s$, is given by:
\begin{equation}
\mathbb{P}_{\mathrm{out}} \triangleq \mathbb{P}(\text{SIR} < \gamma), 
\end{equation}
For a Rayleigh fading channel, the outage probability is given by the following relation \cite{Ref5:Stoc_Geo1}:
\begin{align}
\label{eq:SIR_eq}
\mathbb{P}_{\mathrm{out}} = 1 - e^{-\lambda_{\mathrm{s}} \kappa \pi r^2 \gamma^{\frac {2} {\alpha}}}, 
\end{align}
where $\kappa = \Gamma(1 + \frac {2} {\alpha}) \Gamma(1 - \frac {2} {\alpha})$ with $\Gamma(\cdot)$ being the Gamma function, $\lambda_{\mathrm{s}}$ is the SBS density, $\gamma$ is the target SIR and $r$ is the random distance of the reference user.
\vspace{-0.3cm}
\subsection{Cache Hit Probability}
The cache hit probability is defined as the probability of existence of a given content within $R_{\mathrm{th}}$. The cache hit probability within a threshold distance $R_{\mathrm{th}}$, is represented by $\mathbb{P}(r < R_{\mathrm{th}})$ and is given by:
\begin{equation}
\label{eq:Hit_Prob}
\mathbb{P}(r < R_{\mathrm{th}}) = 1 - e^{-\lambda_{\mathrm{s}} P_c \pi R_{\mathrm{th}}^2},
\end{equation}
This equation ensures the probability of existence of at least one SBS in the coverage area of the user having the requested content. To achieve a given target cache hit probability, the values of $\lambda_{\mathrm{s}}$, $R_{\mathrm{th}}$ and $P_c$ must satisfy the following inequality, given in Lemma \ref{Le_1}:
\begin{lemma}
\label{Le_1}
For a SBS density $\lambda_{\mathrm{s}}$, replication ratio $P_c$ and target cache hit probability $\epsilon$, the following inequality must be satisfied:
\begin{equation}
\label{eq:lea1}
P_c \lambda_{\mathrm{s}} \pi R_{\mathrm{th}}^2 \geq -\ln(1-\epsilon), 
\end{equation} 
\end{lemma}
\vspace{-0.3cm}
\textit{Proof.} Let $\epsilon$ be the target cache hit probability. In order to achieve the target cache hit probability, the right side of eq. \eqref{eq:Hit_Prob} should be greater than or equal to $\epsilon$, i.e., 
\begin{equation*}
\epsilon \leq 1 - e^{-\lambda_{\mathrm{s}} P_c \pi R_{\mathrm{th}}^2}
\end{equation*}
By simplifying the above equation, we get \eqref{eq:lea1}. \\
\indent As there are three variables in Lemma \ref{Le_1}, in order to find the optimal SBS density, we fix some variables to find the bound in the following cases:\\
\textit{\textbf{Case 1}}: For a fixed replication ratio $P_c$, $\lambda_{\mathrm{s}}$ and $R_{\mathrm{th}}$ satisfying eq. \eqref{eq:lea1} are given by the following inequality:
\begin{equation*}
\lambda_{\mathrm{s}} \pi R_{\mathrm{th}}^2 \geq - \frac {\ln (1- \epsilon)} {P_c}, 
\end{equation*}
\textit{\textbf{Case 2}}: As the maximum value of $P_c$ is 1, the lower and upper bound on $P_c$ is given by:
\begin{equation*}
- \frac {\ln(1 - \epsilon)} {\lambda_{\mathrm{s}} \pi R_{\mathrm{th}}^2} \leq P_c \leq 1, 
\end{equation*} 
By assuming that a given user associates with the closest SBS containing content $c$, the probability density function of the random distance $r$, given that the content $c$ exists within $R_{\mathrm{th}}$ is \cite{Ref8:Caching_Ref4}:
\begin{equation}
\label{eq:distribution1}
f_c(r) = 2\pi \lambda_{\mathrm{s}} P_c r \frac {e^{-\lambda_{\mathrm{s}} P_c \pi r^2}} {1 - e^{-\lambda_{\mathrm{s}} P_c \pi R_{\mathrm{th}}^2}}
\end{equation}
\vspace{-0.5cm}
\subsection{Content Outage Probability}
%The Content Outage Probability is defined as the probability of finding at least one replica of the content provided that the SIR does not achieve the SIR threshold ($\gamma$). 
The content outage probability is defined as the probability of not achieving SIR threshold ($\gamma$) for a given content distribution. The content outage probability for a content $c$
%, over a distance $r_s$
, is denoted by $\mathbb{P}_{\mathrm{out}}(c)$. 
\begin{prop}
The content outage probability of content $c$ requested by a reference UE assuming the threshold distance $R_{\mathrm{th}}$, such that Lemma \ref{Le_1} is satified, is given by:
\begin{equation}
\label{eq:Prop1}
\mathbb{P}_{\mathrm{out}}(c) = 1 - \frac {P_c (1 - e^{-\lambda_{\mathrm{s}} (P_c + \kappa \gamma^{\frac {2} {\alpha}} ) \pi R_{\mathrm{th}}^2})} {(1 - e^{-\lambda_{\mathrm{s}} P_c \pi R_{\mathrm{th}}^2})(P_c + \kappa \gamma^{\frac {2} {\alpha}})}
\end{equation}
\end{prop}
\textit{Proof.} In order to find the content outage probability of content, requested by a reference UE, we condition the outage probability of SIR specified in eq. (\ref{eq:SIR_eq}), over the random distance $R = r$ given by:
\begin{equation*}
\begin{split}
\mathbb{P}_{\mathrm{out}}(c) &= \mathbb{E}_R[\mathbb{P}(\mathrm{SIR} < \gamma) | R = r], \\
					 &= \int_{0}^{R_{\mathrm{th}}} \bigg(1 - e^{-{\mathrm{s}} \kappa \pi r^2 \gamma^{\frac {2} {\alpha}}}\bigg ) f_c(r) \mathrm{d}r,
%					 &= \int_{0}^{R} \bigg(1 - e^{-\lambda_b \kappa \pi r^2 \gamma^{\frac {2} {\alpha}}}\bigg ) 2\pi \lambda_b P_c r \frac {e^{-\lambda_b P_c \pi r^2}} {1 - e^{-\lambda_b P_c \pi R^2}} \mathrm{d}r, & \nonumber
\end{split}
\end{equation*}
Plugging \eqref{eq:distribution1} into the above equation and integrating over the threshold distance yields \eqref{eq:Prop1}.
\begin{corollary}
\label{Coro_1}
For a given $\epsilon$, $P_c$ and $R_{\mathrm{th}}$, the optimum number of SBSs is given by:
\begin{equation*}
\lambda_{\mathrm{s}} = - \frac {\ln(1 - \epsilon)} {P_c \pi R_{\mathrm{th}}^2}
\end{equation*}
\end{corollary}
\textit{Proof.} From \eqref{eq:lea1} and \eqref{eq:Prop1}.
\vspace{-0.2cm}
\begin{figure}[t]%
\centering%
\includegraphics*[scale = 0.39]{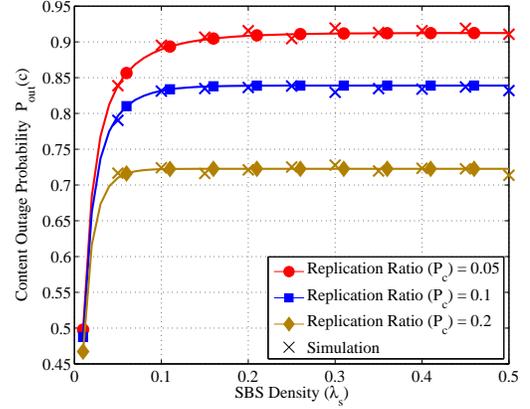}
\caption{Outage probability with respect to SBS density for different replication ratios. $\gamma$ = -10dB, $R_{\mathrm{th}}$ = 5m, $\alpha$ = 3}
\label{fig:Outage_RepRatio_SBSDensity}
\end{figure}%
\section{Numerical Results and Discussion}
\label{sec:Num_Res_Dis}
This section discusses the analytical results obtained in the previous section. We take 5000 realization of $\Phi_{\mathrm{s}}$ to validate the analytical results via simulations. 
%Theoretically, the cache hit probability increases with the distance, according to \eqref{eq:Hit_Prob} at the cost of more interference. If the effect of interference is such that the received SIR is less than the SIR threshold, an outage occurs. 
In what follows, we provide insights on the outage probability in terms of SBS density, replication ratio, SIR threshold and threshold distance. In addition, the simulation results shown in the figures suggest that the analytical model is precise and models the network behavior accurately.
\begin{figure}[b]%
\centering%
\includegraphics*[scale = 0.40]{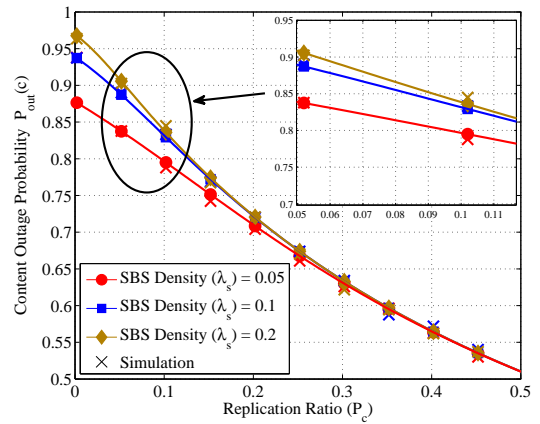}
\caption{Outage probability with respect to Replication Ratio for different SBS densities. $R_{\mathrm{th}}$ = 5m, $\gamma$ = -10dB, $\alpha$ = 3}
\label{fig:Outage_RepRatio_SBSdensity}
\end{figure}%
\vspace{-0.25cm}
\subsection{Impact of SBS Density}
Fig. \ref{fig:Outage_RepRatio_SBSDensity} shows the effect of SBS density on the outage probability for various replication ratios. The figure shows that an increased number of SBSs, at a constant threshold distance, results in increased outage probability due to increased interference. Another implication from the result suggests that the outage probability becomes constant with an increased SBS density. When SBSs cache few contents with a small replication ratio, interference is the dominating factor. Assuming a fixed replication ratio while increasing the SBS density, the effect of interference is increased. Meanwhile, an increased replication ratio, for a fixed SBS density, increases the hit probability because each SBS caches a higher proportion of library contents, which decreases the outage probability.
\begin{figure}[t]%
\centering%
\includegraphics*[scale = 0.385]{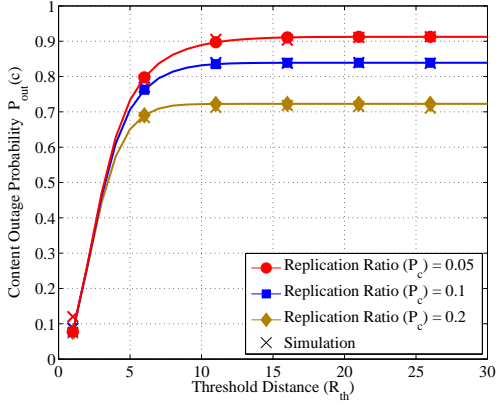}
\caption{Outage probability with respect to distance for different replication ratios. $\gamma$ = -10dB, $\lambda_{\mathrm{s}}$ = $0.1\text{m}^2$, $\alpha$ = 3}
\label{fig:Outage_Distance}
\end{figure}%
\begin{figure}[b]%
\centering%
\includegraphics*[scale = 0.39]{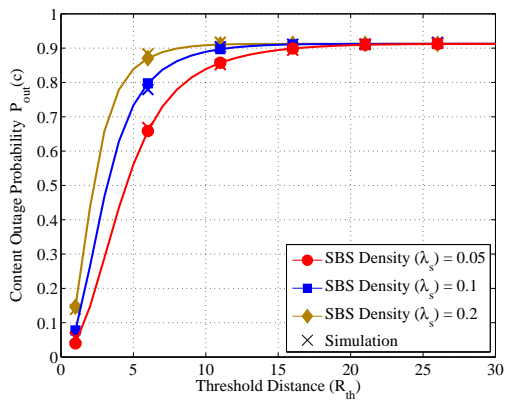}
\caption{Outage probability variation with respect to distance for SBS densities. $P_c$ = 0.02, $\gamma$ = -10dB, $\alpha$ = 3}
\label{fig:Outage_SBSDensity_Distance}
\end{figure}%
\vspace{-0.25cm}
\subsection{Impact of Replication Ratio}
Fig. \ref{fig:Outage_RepRatio_SBSdensity} shows the variation of outage probability as a function of the replication ratio. With an increased replication ratio, the outage probability decreases as each SBS caches a large proportion of the library contents. It can also be seen that for a small SBS density and a small replication ratio, the outage probability is very low. The low value of outage probability is due to the aggregate effect of interference and cache hit probability. Theoretically, a small SBS density with a small replication ration decreases the hit probability. As the interference effect is very small at a small SBS density, the cache hit probability is the dominating factor. 
%Meanwhile, it can also be deduced from the results that the convergence rate increases with the SBS density as the effect of interference remains the same while increasing the replication ratio. %Another implication from the figure suggests that the outage probability coverages to 0.05 irrespective of SBS density.
\begin{figure}[t]%
\centering%
\includegraphics*[scale = 0.405]{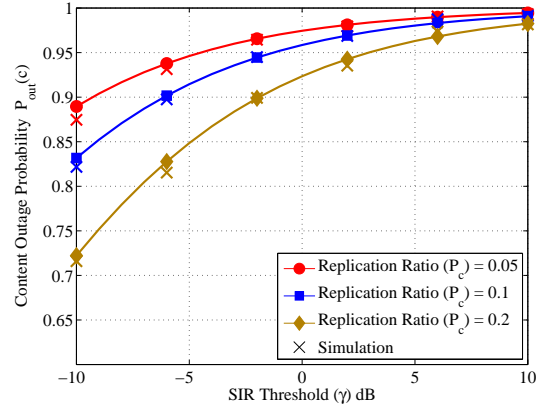}
\caption{Evolution of Outage probability with respect to SIR threshold for different replication ratio. $\lambda_{\mathrm{s}}$ = 0.1, $R_{\mathrm{th}}$ = 10m, $\alpha$ = 3}
\label{fig:Outage_RepRatio_SIRThreshold}
\end{figure}%
\begin{figure}[b]%
\centering%
\includegraphics*[scale = 0.38]{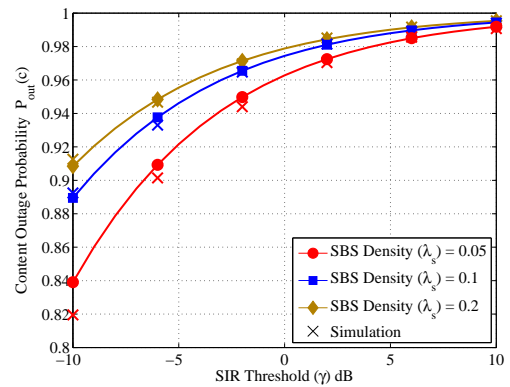}
\caption{Evolution of Outage probability with respect to SIR threshold for different SBS densities. $P_c$ = 0.02, $R_{\mathrm{th}}$ = 10m, $\alpha$ = 3}
\label{fig:Outage_SIRThreshold_SBSdensity}
\end{figure}%
\vspace{-0.25cm}
\subsection{Impact of Distance}
Fig. \ref{fig:Outage_Distance} shows the effect of coverage radius $(R_{\mathrm{th}})$ on the outage probability for various replication ratios. This demonstrates that the outage probability increases with the threshold distance. If the threshold distance increases, the number of SBSs increases and the hit probability for the content increases. However, with an increased distance, the link reliability decreases due to increased interference. Another insight drawn from the figure is that with increased replication ratio, the outage probability decreases as every SBS caches most of the library contents. In addition, the outage probability levels off, as the threshold distance is increased beyond 10m. This is due to the fact that as the distance is increased, the effect of interference becomes negligible. Fig. \ref{fig:Outage_SBSDensity_Distance} shows the variation of outage probability with distance for different SBS density. As the threshold distance is increased, the outage probability increases due to the higher interference level. Further increasing the distance improves the hit probability at the cost of more interference, thereby impacting the outage probability. Besides, with the increased SBS density, the outage probability increases due to increased interference from the SBSs.
\begin{figure}[t]%
\centering%
\includegraphics*[scale = 0.38]{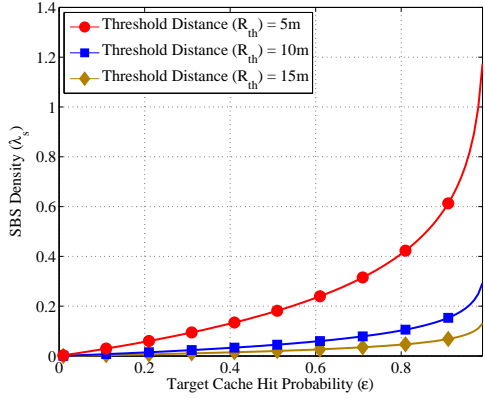}
\caption{Optimal SBS density for different distances and target hit probability. $P_c$ = 0.1}
\label{fig:Opt_SBS_distance}
\end{figure}%WWW
\begin{figure}[b]%
\centering%
\includegraphics*[scale = 0.385]{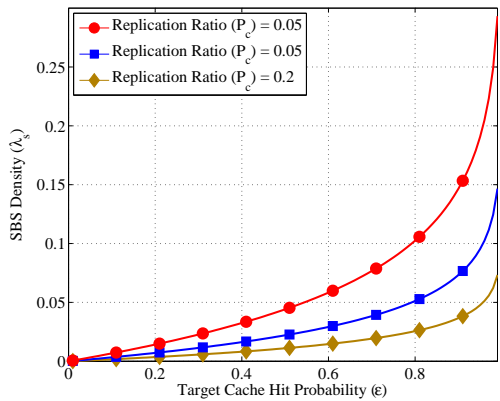}
\caption{Optimal SBS density for different replication ratio and target hit probability. $R_{\mathrm{th}}$ = 10m}
\label{fig:Opt_SBS_Repratio}
\end{figure}%
\vspace{-0.2cm}
\subsection{Impact of SIR Threshold}
Fig. \ref{fig:Outage_RepRatio_SIRThreshold} presents the variation of outage probability with respect to SIR threshold for different replication ratios. It can be seen that with the increased SIR threshold, the outage probability increases. The increased SIR threshold relaxes the interference requirement, resulting in an increased outage. Meanwhile, the outage converges to 1 as the SIR threshold increases. However, the convergence is faster for small replication ratio. The fast convergence for small replication is due to the fact that the effect of interference is dominating than the caches of SBSs. \\
\indent Fig. \ref{fig:Outage_SIRThreshold_SBSdensity} shows the variation of outage probability with SIR threshold for different SBS density. It can be seen that the outage probability increases with SIR threshold as the SBS density is increased. The increased SIR threshold lessens the interference requirement, resulting in an increased outage. In addition, the outage probability coverges to 1 irrespective of the SBS density. However, the convergence rate for small SBS density is less than the convergence rate for large SBS density. For small SBS density, the aggregate interference of SBSs is much smaller instead of large SBSs density.
\vspace{-0.25cm}
\subsection{Optimum SBS Density}
Fig. \ref{fig:Opt_SBS_distance} shows the optimum number of SBSs as a function of target hit probability for a given $R_{\mathrm{th}}$ and $P_c$, given by Corollary \ref{Coro_1}. The first implication, from the figure, suggests that an increased threshold distance requires a small SBS density for a given replication ratio. With the increased threshold distance, a large number of SBSs becomes available in the coverage area of UE, thus, requiring small SBS density. However, the exponential behavior of the curve for large distance suggests that it is quite challenging to achieve a high target cache hit probability for such large threshold distance. In addition, the figure reveals that with an increased target cache hit probability, the SBS density for different distances varies slowly and the optimum SBS density converges. Fig. \ref{fig:Opt_SBS_Repratio} shows the optimum number of SBSs required to achieve a target hit probability at a fixed distance for different replication ratios. It can be seen that a high replication ratio decreases the number of SBSs as many SBSs caches most of the library contents. Meanwhile, it is difficult to obtain a highest target cache hit probability for a small replication ratio. Therefore, small distances and replication ratios require a high SBS density for a given target cache hit probability. Moreover, the figure reveals that as the replication ratio is increased, the variation in the SBS density becomes smaller to achieve a target cache hit probability.   
\vspace{-0.2cm}
\section{Conclusion and Future Work}
In this paper, we studied a cache enabled small cell network comprising of SBSs that store contents from a library. By considering the distribution of SBSs to be a PPP, we derived the outage probability of getting the requested content over a threshold distance. In addition, we characterized the optimum number of SBSs to achieve a target hit probability. Finally, we performed numerical analysis to show the interplay between outage probability, SBS density, threshold distance and replication ratio. Future works involve incorporating the modeling of wired and wireless backhauling, and investigating other network deployments such as clustered PPPs.  
%%\item Average Loss Rate per content as a function of distance (R)
%\begin{figure}[t]%
%\centering%
%\includegraphics[width = 80mm,height = 60mm]{HitProb_StorageSize}
%\caption{Hit Probability as a function of Cache Size for different SBS Density}
%\label{fig:HitProb_StorageSize}
%\end{figure}%
\vspace{-0.2cm}

% conference papers do not normally have an appendix

% use section* for acknowledgement
%\section*{Acknowledgment}
%
%
%The authors would like to thank...

% trigger a \newpage just before the given reference
% number - used to balance the columns on the last page
% adjust value as needed - may need to be readjusted if
% the document is modified later
%\IEEEtriggeratref{8}
% The "triggered" command can be changed if desired:
%\IEEEtriggercmd{\enlargethispage{-5in}}

% references section

% can use a bibliography generated by BibTeX as a .bbl file
% BibTeX documentation can be easily obtained at:
% http://www.ctan.org/tex-archive/biblio/bibtex/contrib/doc/
% The IEEEtran BibTeX style support page is at:
% http://www.michaelshell.org/tex/ieeetran/bibtex/
%\bibliographystyle{IEEEtran}
% argument is your BibTeX string definitions and bibliography database(s)
%\bibliography{IEEEabrv,../bib/paper}
%
% <OR> manually copy in the resultant .bbl file
% set second argument of \begin to the number of references
% (used to reserve space for the reference number labels box)

% that's all folks
\end{document}